# Progress on Perfect Lattice Actions for QCD

W. Bietenholz[a], R. Brower[a b], S. Chandrasekharan[a] and U.-J. Wiese[a] *

[a]CTP-LNS, Massachusetts Institute of Technology, Cambridge MA 02139, USA

[b]Dept. of Physics, Boston Univ., 590 Commonwealth Ave., Boston MA 02215, USA

We describe a number of aspects in our attempt to construct an approximately perfect lattice action for QCD. Free quarks are made optimally local on the whole renormalized trajectory and their couplings are then truncated by imposing 3-periodicity. The spectra of these short ranged fermions are excellent approximations to continuum spectra. The same is true for free gluons. We evaluate the corresponding perfect quark-gluon vertex function, identifying in particular the "perfect clover term". First simulations for heavy quarks show that the mass is strongly renormalized, but again the renormalized theory agrees very well with continuum physics. Furthermore we describe the multigrid formulation for the non-perturbative perfect action and we present the concept of an exactly (quantum) perfect topological charge on the lattice.

A large number of contributions to this conference are devoted to improved actions; there is no doubt that they are in fashion. This indicates a consensus that they represent a ray of hope for a great leap forward in lattice QCD. Most improvement procedures follow in one way or the other Symanzik's program [1], using perturbation theory in the lattice spacing, e.g. [2]. Our work employs a different concept, utilizing renormalization group tools to construct an approximately perfect action. A fixed point action (FPA) on a critical surface is an example of a perfect action, an action without any cutoff artifacts. Using the fixed point action even at finite correlation length, a drastically improved scaling behavior has been observed for the 2d O(3) model [3] and pure 4d SU(3) gauge theory [4], and is expected also for other asymptotically free theories such as full QCD.

## 1. Free fermions

Fixed point actions have been constructed also for free fermions [5,6]. They can be obtained from iterating block renormalization group transformations (RGTs) with a finite blocking factor, or more efficiently by sending the blocking factor to infinity and performing only one step. This amounts to a technique that we call "blocking from the continuum". It has been used extensively in a discussion of the Schwinger model [7] and for quarks and gluons [8]. One starts from a continuum theory, divides the coordinate space into lattice cells and defines lattice fields by integrating over these cells. For free fermions the lattice action is then given by

$$e^{-S[\bar{\Psi},\Psi]} = \int D\bar{\psi} D\psi \exp\Big\{ - s[\bar{\psi},\psi] \qquad (1)$$
$$-\frac{1}{a}\sum_x \Big[\bar{\Psi}_x - \int_{c_x} \bar{\psi}(y)dy\Big]\Big[\Psi_x - \int_{c_x} \psi(y)dy\Big]\Big\}.$$

Here $\bar{\psi}, \psi$ are continuum fields, $s$ is the continuum action and $c_x$ is a unit hypercube with center $x$. $S$ is the perfect action in terms of the lattice fields $\bar{\Psi}, \Psi$. Finally $a \geq 0$ is an arbitrary RGT parameter; for any choice of $a$ all expectation values are invariant under this RGT. For $a \to 0$ this is a $\delta$ function RGT, and $a > 0$ "smears" the $\delta$ function to a Gaussian.

For fermions of mass $m$, this yields in momentum space the perfect action

$$S[\bar{\Psi},\Psi] = \frac{1}{(2\pi)^d}\int_B d^d p \bar{\Psi}(-p)\Delta^f(p)^{-1}\Psi(p)$$

*Based on two talks presented by W.B. and S.C. and two posters presented by R.B. and U.-J.W. Work supported by U.S. Department of Energy (D.O.E.) under cooperative research agreement DE-FC02-94ER40818.

$$\begin{aligned}\Delta^f(p) &= \sum_{l\in \mathbb{Z}^d} \frac{\Pi(p+2\pi l)^2}{i(\not p + 2\pi \not l) + m} + a \\ \Pi(p) &= \prod_{\nu=1}^{d} \frac{\hat{p}_\nu}{p_\nu}, \qquad \hat{p}_\nu = 2\sin(p_\nu/2). \qquad (2)\end{aligned}$$

Here $B = ]-\pi,\pi]^d$ is the Brillouin zone, $\Delta^f$ is the perfect lattice propagator and the function $\Pi$ is the Fourier transform of the step function which is 1 in $c_0$ and 0 otherwise. This construction is illustrated in Fig. 1. The blocking attaches a $\Pi$ function to $\bar{\Psi}$ and to $\Psi$, and the points in the blocks are connected by the continuum propagator. The sum over $l$ probes all points inside the blocks. The $\Pi$ functions assure its convergence.

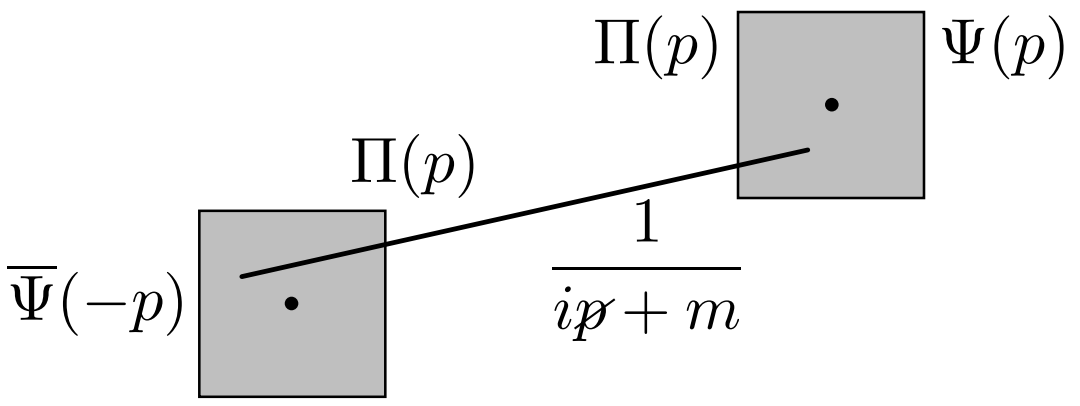

Figure 1. Construction of the perfect propagator for free fermions.

For $m = 0$ this is a fixed point action, and for general $m$ we have an entire renormalized trajectory for free fermions. We can go back to coordinate space and write the perfect action as

$$S[\bar{\Psi},\Psi] = \sum_{x,y} \bar{\Psi}_x [i\rho_\mu(x-y)\gamma_\mu + \lambda(x-y)]\Psi_y . \quad (3)$$

It turns out that for any $a > 0$ this action is local in the sense that $\rho_\mu$ and $\lambda$ decay exponentially [5]. We can now tune $a$ to optimize the locality. In $d = 1$ the summation in $\Delta^f(p)$ can be done analytically and the action has only nearest neighbor couplings iff

$$a(m) = \frac{e^m - m - 1}{m^2}. \quad (4)$$

In higher dimensions we cannot achieve such "ultralocality" (i.e. a finite support for $\rho_\mu$ and $\lambda$), but the above choice for $a$ still provides an extremely local perfect action, see Figs. 2 and 3. These figures show that we can also handle heavy quarks, where the action becomes even more local.

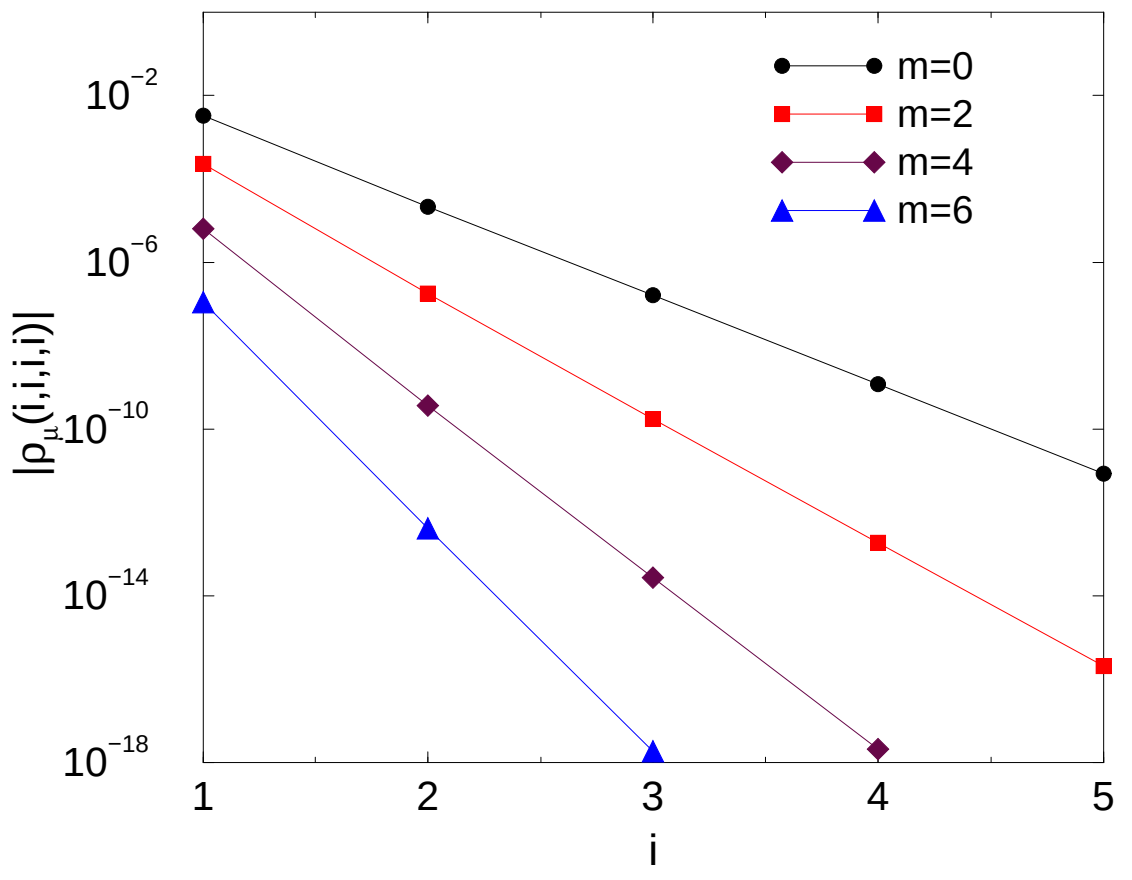

Figure 2. The exponential decay of $|\rho_\mu|$ on the 4-space diagonal.

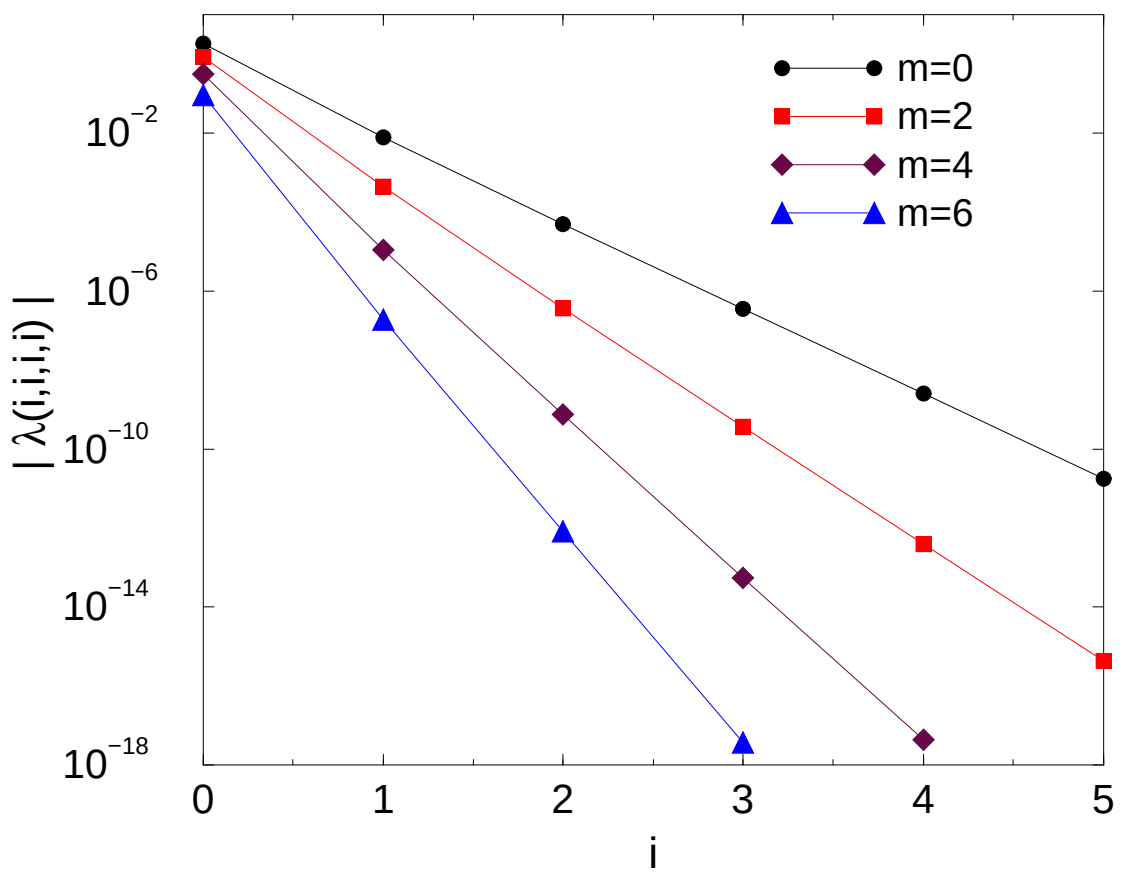

Figure 3. The exponential decay of $|\lambda|$ on the 4-space diagonal.

As an application to thermodynamics, we consider the following ratio for free massless fermions

$$\frac{p}{T^4} = \frac{7\pi^2}{180} = 0.3838\ldots, \quad \text{(continuum)} \quad (5)$$

where $p$ is the pressure and $T$ the temperature. Fig. 4 shows this ratio at $N_t$ lattice points in the 4 direction for Wilson fermions and the above fixed point fermion. For the latter the cutoff artifacts are strongly suppressed; the remaining artifacts at very small $N_t$ correspond to temperature dependent prefactors that we have neglected in the partition function when we preformed the RGT.

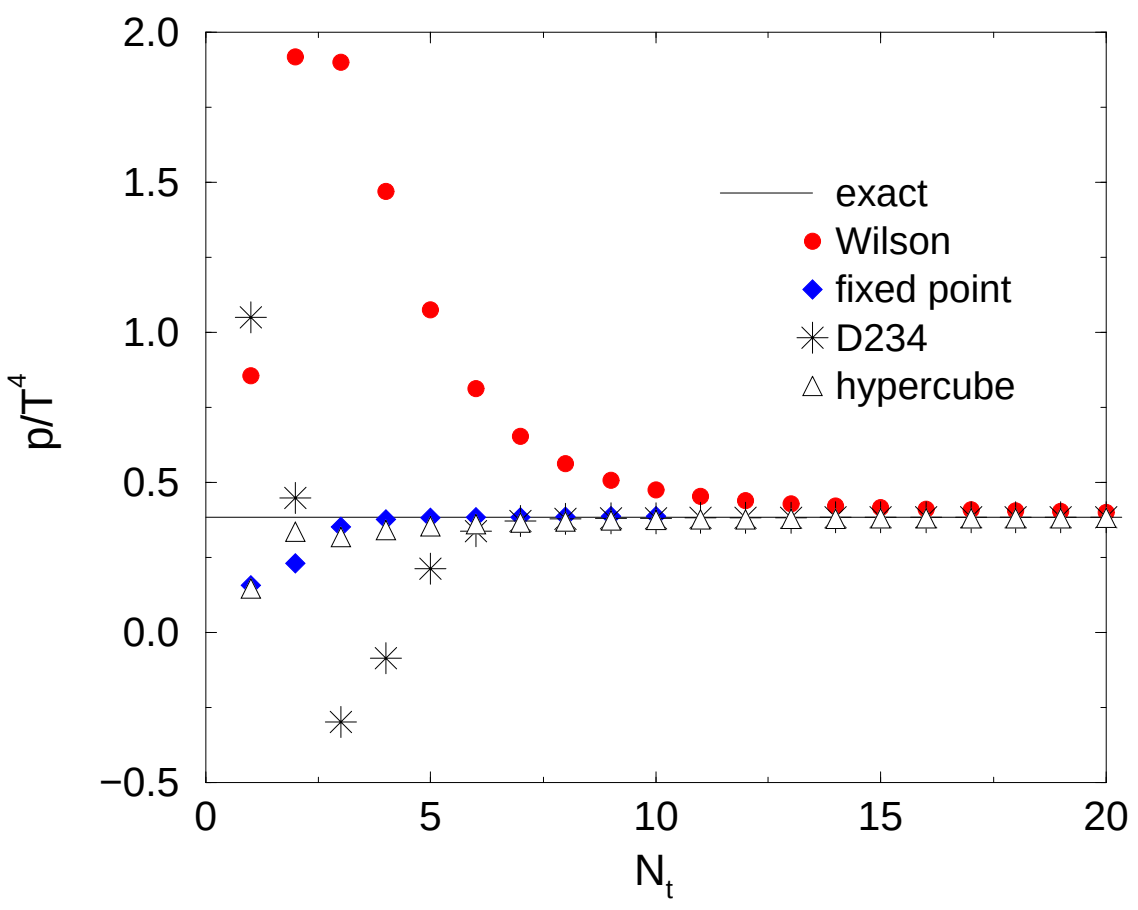

Figure 4. The ratio $p/T^4$ versus $N_t$ for free massless Wilson, fixed point and hypercube fermions.

| coupling | $m=0$ | $m=2$ |
|---|---|---|
| $\rho_1(1000)$ | 0.136846794 | 0.0185415007 |
| $\rho_1(1100)$ | 0.032077284 | 0.0031625467 |
| $\rho_1(1110)$ | 0.011058131 | 0.0007898101 |
| $\rho_1(1111)$ | 0.004748991 | 0.0002501304 |
| $\lambda(0000)$ | 1.852720547 | 0.8442376349 |
| $\lambda(1000)$ | −0.060757866 | −0.0119736477 |
| $\lambda(1100)$ | −0.030036032 | −0.0032647950 |
| $\lambda(1110)$ | −0.015967620 | −0.0011445684 |
| $\lambda(1111)$ | −0.008426812 | −0.0004622883 |

Table 1
Couplings for a free hypercube fermion action, which are perfect for 3-periodic configurations. Note that all coordinates of $\lambda$ and all non-$\mu$ coordinates of $\rho_\mu$ can be permuted and sign flipped arbitrarily. $\rho_\mu$ is odd in the $\mu$ coordinate.

To develop a practical scheme one must truncate the exponentially small tails in perfect actions at a finite range. Here we consider a *hypercube action* for quasi-perfect free fermions. Instead of simply setting the couplings outside a unit hypercube to zero, we start from the observation that one can easily obtain perfect actions in a finite volume with periodic boundary conditions by reducing the integral $\int_B d^dp$ in Eq. (2) to a discrete sum. In particular if we use 3-periodic boundary conditions, we obtain couplings for the free action that naturally live within a hypercube and the resultant hypercube Dirac operator is determined by nine couplings, which are given for two masses in Table 1. Using these couplings in the infinite volume represents an elegant truncation scheme, which maintains the exact result in momentum space for $p_\mu = 0, \pm 2\pi/3$. The ratio $p/T^4$ for this hypercube fermion (at $m=0$) is also shown in Fig. 4.

The perfect action reproduces the continuum spectrum exactly. Truncation to a hypercube distorts the spectrum a little. However, it is still very good. In Figs. 5 and 6 we plot the dispersion relation of the hypercube action for the massless and a massive case, respectively. We compare it to the standard Wilson fermion action, the D234 action [9] (see also Fig. 4) and a recent proposal by the Fermilab group [10].

There is a major difference between the on-shell Symanzik improvement and the perfect action improvement after truncation. The on-shell improvement concentrates on obtaining the spectrum correctly order by order in the lattice spacing (i.e., small momenta). On the other hand, in the truncated perfect actions, there are errors at each order of the lattice spacing. However, these errors are exponentially small in the range of the truncation so that the dispersion relation is good over nearly the *entire* Brillouin zone, as Fig. 5 and 6 show. Still one may analyze specifically the errors for small momenta. For example in the case of massless free fermions the small momentum expansion of the Dirac operator is given by

$$\Delta^f(p)^{-1} \cong i\gamma_\mu p_\mu[1 + c_1 p_\mu^2 + c_2 p^2] + c_3\, p^2. \qquad (6)$$

The breaking of rotational invariance depends on $c_1$. In the Symanzik on-shell improvement, $c_1$ is tuned to zero. For the Wilson fermion we obtain $c_1 \cong -0.166$ and for the hypercube action $c_1 \cong 0.024$. In the case of massive fermions the dispersion relation can be expanded in small momenta: $E = m_s + p^2/2m_k + \dots$, where $m_s$ is referred to as the *static mass*, and $m_k$ as the *kinetic mass*. Ideally one has $m_s = m_k$. In the Symanzik approach this can again be tuned. In the truncated perfect action $m_k$ is close to $m_s$ but not exactly equal. In Fig. 7 we plot $m_k$ as a function of $m_s$ for the hypercube action, and compare it with the Wilson action. The fact that $m_k$ is a little larger than $m_s$ for the hypercube fermion can also be seen in Fig. 6, where the curvature at $p = 0$ is slightly too large. Tuning $m_k$ to the exact value improves the dispersion modestly at $p << 1$, but can make it much worse at $p = O(1)$, as the Fermilab curve shows.

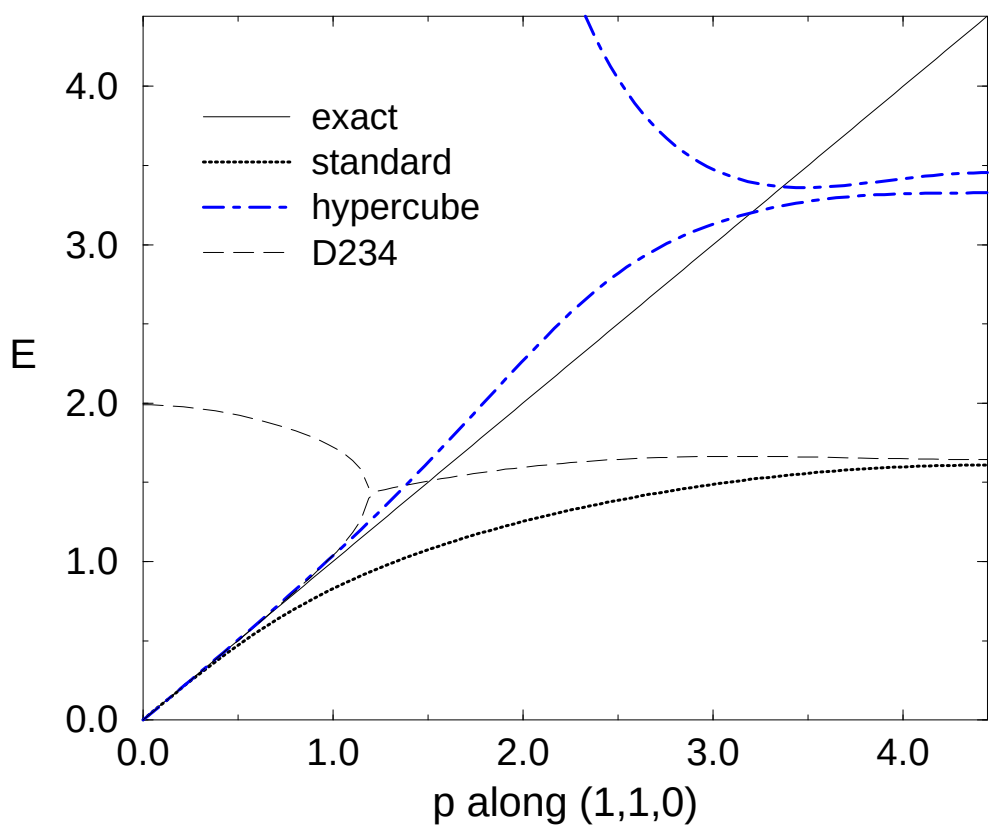

Figure 5. Comparison of dispersion relations at $m = 0$.

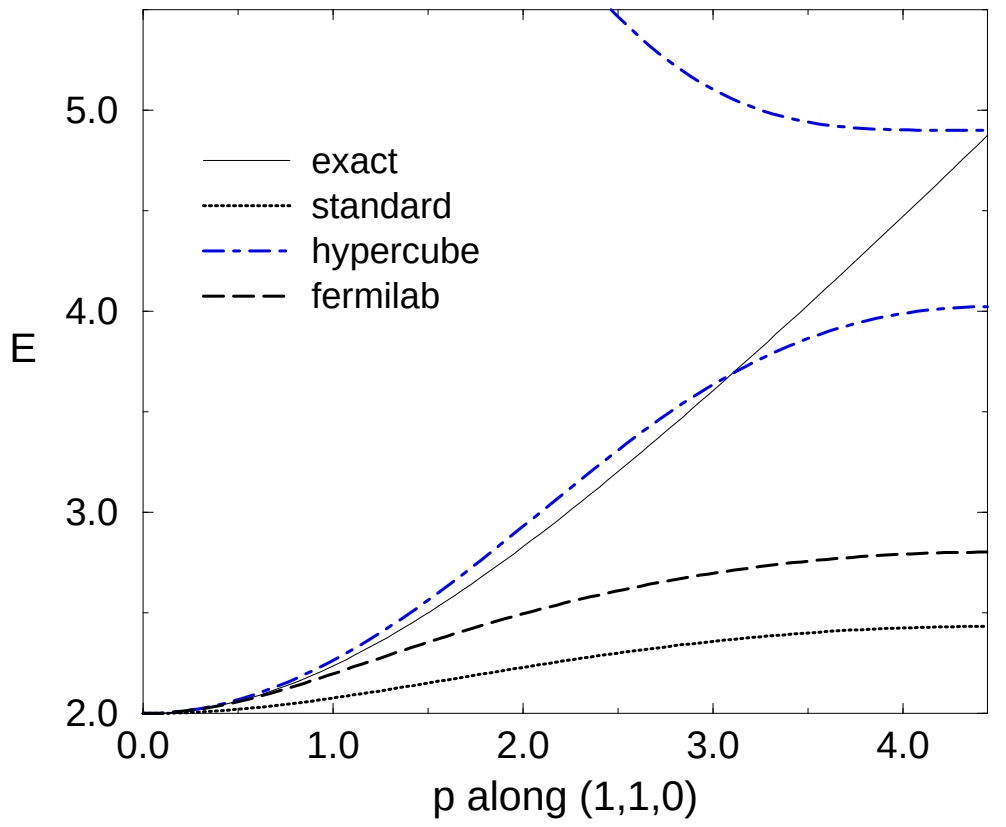

Figure 6. Comparison of dispersion relations at $m = 2$.

The above FPA was obtained using an RGT which manifestly breaks chiral symmetry. Consequently, the lattice fermions resemble the Wilson fermions in their chiral properties. In fact, iterating RGTs on the lattice, we obtain the above FPA if we start from Wilson fermions. However, perfect actions with the properties of *staggered fermions* are also of interest. In particular the remnant U(1)⊗U(1) chiral symmetry simplifies the study of chiral symmetry breaking. In Ref. [11] it was shown how one can block staggered fermions from a fine to a coarse lattice (with blocking factor 3) without mixing the pseudoflavors. Later this blocking prescription was used to construct a perfect action for free staggered fermions [6,12]. The same results can be derived using the technique of “blocking from the continuum” [13].

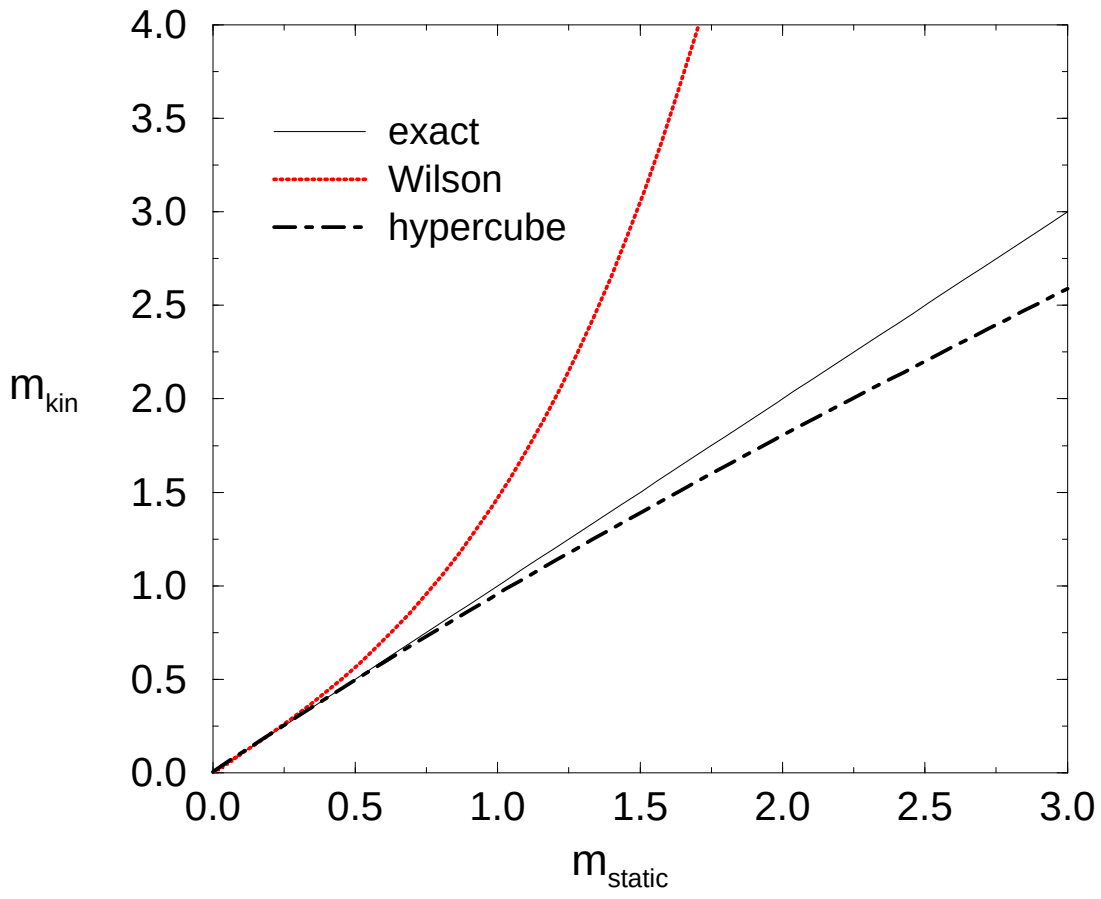

Figure 7. Static vs. kinetic mass.

The perfect action for free staggered fermions can be written in momentum space as

$$S[\bar{\chi},\chi] = \frac{1}{(2\pi)^d}\int_B d^d p\ \bar{\chi}_\rho(-p)\Delta^f_{\rho\rho'}(p)^{-1}\chi_{\rho'}(p),$$
$$\Delta^f_{\rho\rho'}(p) = -i\alpha_\mu(p)\ \Gamma^\mu_{\rho\rho'}(p) + \beta(p)\ \delta_{\rho,\rho'}. \qquad (7)$$

The indices $\rho, \rho'$ run over the sixteen pseudoflavors. A unitary transformation connects this pseudoflavor basis to the Dirac ⊗ flavor basis. The matrix $\Gamma^\mu_{\rho\rho'}(p)$ is unitarily equivalent to $(\gamma_\mu \otimes \mathbf{1})$, and is given by $\eta_\mu(\rho)[\delta_{\rho-\hat{\mu},\rho'} + \delta_{\rho+\hat{\mu},\rho'}]e^{ip(\rho-\rho')/2}$, with the sign factor $\eta_\mu(\rho) = (-1)^{\sum_{\nu<\mu}\rho_\nu}$. The functions $\alpha_\mu(p)$ and $\beta(p)$ take the form

$$\alpha_\mu(p) = \sum_{l\in \mathbb{Z}^d} \frac{(p_\mu + 2\pi l_\mu)\Pi(p+2\pi l)^2}{(p+2\pi l)^2 + m^2}(-1)^{l_\mu} + c\hat{p}_\mu,$$
$$\beta(p) = \sum_{l\in \mathbb{Z}^d} \frac{m}{(p+2\pi l)^2 + m^2}\Pi(p+2\pi l)^2 + a.$$

The RGT parameters $c$ and $a$ are used to optimize locality of the action. $a$ is a mass-like smearing parameter of the $\delta$ function analogous to the parameter $a$ used before, and $c$ is its kinetic counterpart.

In the above representation the pseudoflavors are placed on a lattice with unit spacing. When gauge fields are introduced, it will be convenient to place the fermions on a finer lattice with spacing 1/2, distributing the pseudoflavors over the 16 corners of the hypercube of the finer lattice. Then there is one variable per site of the finer lattice and the action takes the form

$$S[\bar{\chi},\chi] = \sum_{x,y} \bar{\chi}_x [i\eta_\mu(x-y)\rho_\mu(x-y) + \lambda(x-y)]\chi_y \tag{8}$$

where $\rho_\mu(x-y)$ is non-zero only when $(x-y)_\mu \in \{\pm 1/2, \pm 3/2, \ldots\}$ and $(x-y)_{\nu\neq\mu}$ is an integer, and $\lambda(x-y)$ is non-zero only when $(x-y)_\nu$ is an integer for all $\nu$. Due to the symmetries involved, $\rho_1(x-y)$ and $\lambda(x-y)$ contain all the information.

Again in $d=1$ the summation over $l$ can be done analytically and the action is ultralocal iff

$$a(m) = \frac{\sinh(m) - m}{m^2}, \quad c(m) = \frac{\cosh(m/2) - 1}{m^2}.$$

Following the example of Wilson fixed point fermions, we use the same parameters in higher dimensions for optimal locality. In Table 2 we give the largest couplings for mass $m=0$ and $m=2$.

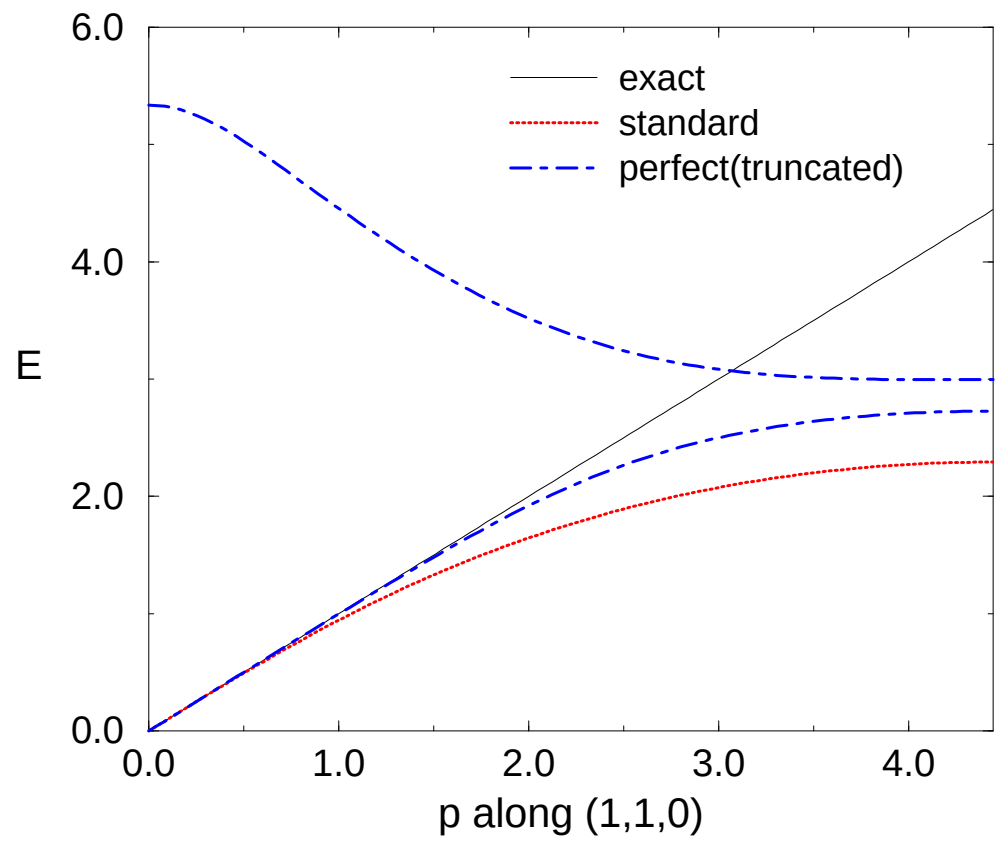

Figure 8. Dispersion relations for staggered fermions at $m=0$.

Using the couplings for $m=0$ and omitting the distance 2 couplings, we obtain the dispersion relation for the free staggered fermions shown in Fig. 8.

| $(x,y,z,t)$ | $\rho_1(x+0.5,y,z,t)$ | $\lambda(x,y,z,t)$ |
|---|---|---|
| m=0.0 | | |
| (0000) | 0.6617391 | 0.0 |
| (0100) | 0.0441181 | 0.0 |
| (0110) | 0.0046569 | 0.0 |
| (0111) | 0.0004839 | 0.0 |
| (1000) | 0.0234887 | 0.0 |
| (1100) | −0.0004933 | 0.0 |
| (1110) | −0.0009913 | 0.0 |
| (1111) | −0.0004819 | 0.0 |
| (0200) | 0.0018423 | 0.0 |
| (0210) | 0.0001419 | 0.0 |
| (1200) | −0.0001211 | 0.0 |
| (1210) | −0.0001011 | 0.0 |
| m=2.0 | | |
| (0000) | 0.3586038 | 0.9799873 |
| (0100) | 0.0154803 | 0.0246899 |
| (0110) | 0.0012088 | −0.0015531 |
| (0111) | 0.0001282 | −0.0009711 |
| (1000) | 0.0071677 | 0.0246899 |
| (1100) | −0.0004502 | −0.0015531 |
| (1110) | −0.0002537 | −0.0009711 |
| (1111) | −0.0000832 | −0.0003371 |
| (0200) | 0.0003688 | 0.0005581 |
| (0210) | 0.0000096 | −0.0000703 |
| (1200) | −0.0000366 | −0.0000703 |
| (1210) | −0.0000155 | −0.0000265 |

Table 2
The largest values of $\rho_1(r)$ and $\lambda(r)$ for free staggered fermions at mass $m=0$ and $m=2$.

## 2. Free gauge fields

Since gauge fields live naturally on the lattice links, the blocking scheme for free gauge fields is different from matter fields. An adequate prescription for their blocking from the continuum was given in [7,8]. One integrates over all straight connections of corresponding points in adjacent cells. Hence one identifies the lattice gauge field (living on link centers)

$$A_\mu(p) = \sum_l a_\mu(p+2\pi l)\Pi_\mu(p+2\pi l)(-1)^{l_\mu}$$

$$\Pi_\mu(p) = \frac{\hat{p}_\mu}{p_\mu}\Pi(p) \qquad (9)$$

where $a_\mu$ is a continuum gauge field. In analogy to Eq. (1) one finds an FPA for free gauge fields. In a general gauge it reads

$$S[A_\mu] = \frac{1}{(2\pi)^d}\int_B dp A_\mu(-p)\Delta^g_{\mu\nu}(p)^{-1}A_\nu(p)$$
$$\Delta^g_{\mu\nu}(p)^{-1} = \Delta^g_\mu(p)^{-1}\delta_{\mu\nu} - \frac{\hat{p}_\mu\Delta^g_\mu(p)^{-1}\Delta^g_\nu(p)^{-1}\hat{p}_\nu}{\sum_\rho \hat{p}_\rho^2\Delta^g_\rho(p)^{-1}}$$
$$\Delta^g_\mu(p) = \sum_l \frac{\Pi_\mu(p+2\pi l)^2}{(p+2\pi l)^2} + \alpha + \gamma\hat{p}_\mu^2. \qquad (10)$$

$\alpha$ and $\gamma$ are RGT parameters analogous to $a$ and $c$ in the case of staggered fermions. We tune both of them to optimize locality. The FPA for $\alpha = 1/6$, $\gamma = -1/72$ is in $d = 2$ the standard plaquette action, and in four dimensions it is still extremely local [8]. The perfect free gluon can be truncated to a unit hypercube in the same way it was done for fermions, and the dispersion relation for transverse gluons is still excellent, see Fig. 9.

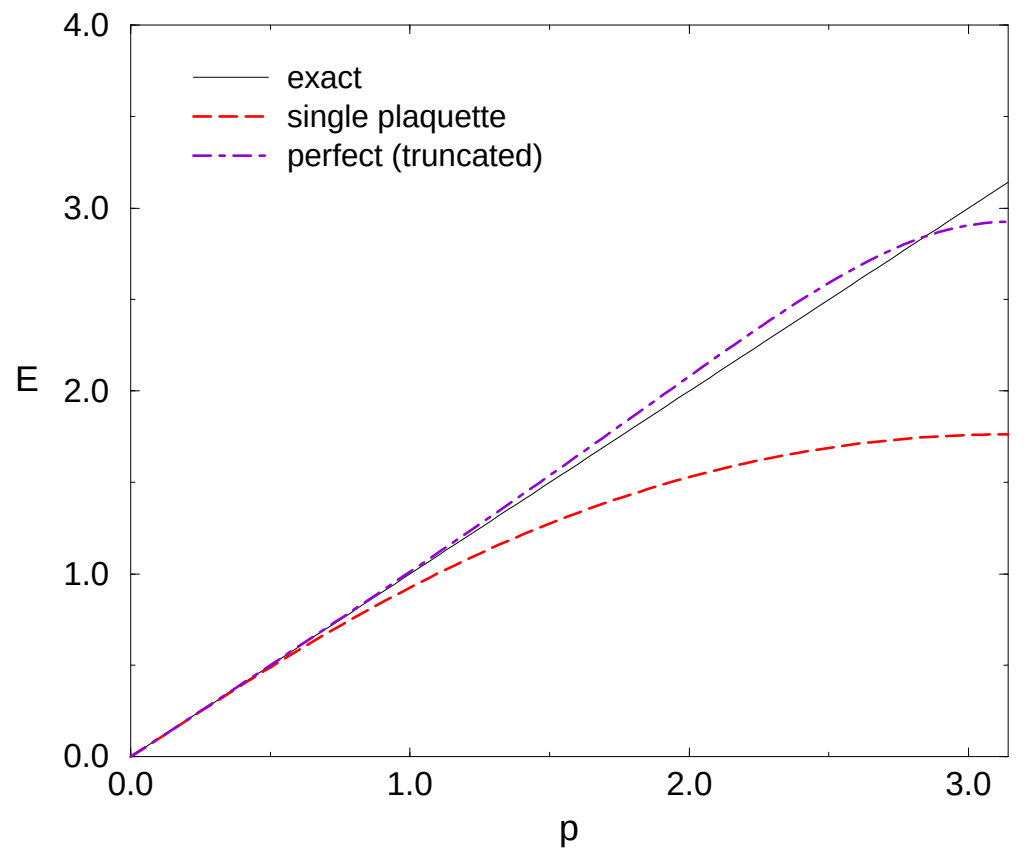

Figure 9. Dispersion relation of the transverse gluon.

To impose gauge invariance one can parameterize this weak coupling FPA in terms of closed loops. In Sec 6. we suggest a 6 loop parameterization of the gauge action (see Eq. 26), which is a good approximations for weak fields and which can therefore be used to initialize our multigrid minimization to obtain non-perturbative results.

As we noted above the *staggered fermions* require an RGT that distinguishes the pseudoflavors. For compatibility, the blocking convolution of the gauge field also has to be altered. The reason is that in the interacting theory the gauge field has to couple to the pseudoflavors appropriately. This can be achieved by blocking the gauge field from the continuum in the following way

$$A_{\mu,x} = \int d^d y M_\mu(y) a_\mu(x-y), \qquad (11)$$
$$M_\mu(y) = \begin{cases} M(y_\mu) & |y_\nu| \le \frac{1}{2}, \nu \neq \mu \\ 0 & \text{otherwise} \end{cases},$$
$$M(y_\mu) = \begin{cases} 1 & |y_\mu| \le \frac{1}{4} \\ \frac{3}{2} - 2|y_\mu| & \frac{1}{4} \le |y_\mu| \le \frac{3}{4} \\ 0 & \text{otherwise} \end{cases}.$$

Because of its shape we call $M_\mu$ the "Mansard" function. In momentum space the expression analogous to (9) reads

$$A_\mu(p) = \sum_{l\in \mathbb{Z}^d} a_\mu(p+4\pi l)\Pi^M_\mu(p+4\pi l)(-1)^{l_\mu}$$
$$\Pi^M_\mu(p) = \frac{4\sin(p_\mu/4)}{p_\mu}\Pi(p). \qquad (12)$$

$A_\mu(p)$ is $4\pi$ antiperiodic, because it lives on the link centers of a lattice with spacing 1/2. We just replace the function $\Pi_\mu$ by $\Pi^M_\mu$, which is the Fourier transform of the Mansard function, and the momentum $p$ by $p/2$. The Brillouin zone $B$ then refers to $]2\pi, 2\pi]$. Also the fixed point propagator maintains the form given in Eq. (10), up to the same substitutions (and $2\pi l$ replaced by $4\pi l$). Ultralocality in $d = 2$ and optimal locality in $d = 4$ now requires the momentum dependent smearing coefficients

$$\alpha(p) = \frac{1}{8}\Big(\frac{1}{2} + \frac{1}{3}\prod_\mu \cos^2\frac{p_\mu}{4}\Big),$$
$$\gamma(p) = -\frac{1}{32}\Big(\frac{1}{4} + \frac{1}{9}\prod_\mu \cos^2\frac{p_\mu}{4}\Big). \qquad (13)$$

Note that $\alpha + \gamma\hat{p}_\mu^2$ is always positive, so that the RGT is well defined.

## 3. Interacting theory

If we introduce finite interactions, it is no longer possible to block the fields analytically.

One can either use perturbation theory or numerical RGTs. However, one may hope for some improvement in non-perturbative physics by using just the (truncated) FPA of the free fermions. A simple test is the mesonic dispersion relation after rendering the fermionic action gauge invariant by hand. In order to study the mesonic dispersion relation using the hypercube action, a quenched simulation with the Wilson gauge action was performed at $\beta = 5.0$ on a $6^3 \times 18$ lattice. We used the $m = 0$ couplings of Table 1 and we obtained a static meson mass of $M = 3.0$. Thus, there is a huge renormalization of the quark mass and the simulation is only relevant to heavy quark physics. Since these simulations did not use the perfect quark-gluon interactions this is acceptable. In heavy quark physics the kinetic mass of the meson plays an important role. If it is equal to the static mass $M$, one obtains for the effective speed of light $\sqrt{E^2 - M^2}/p = 1$ for small momenta. In the continuum this condition is imposed by Lorentz invariance and it holds for arbitrary momenta. It can be used to probe the behavior of a lattice theory. In Fig. 10 we plot the result of simulations for heavy mesons. One sees a dramatic improvement for the hypercube fermions compared to Wilson fermions. This result suggests that renormalization preserves Lorentz invariance in the truncated theory.

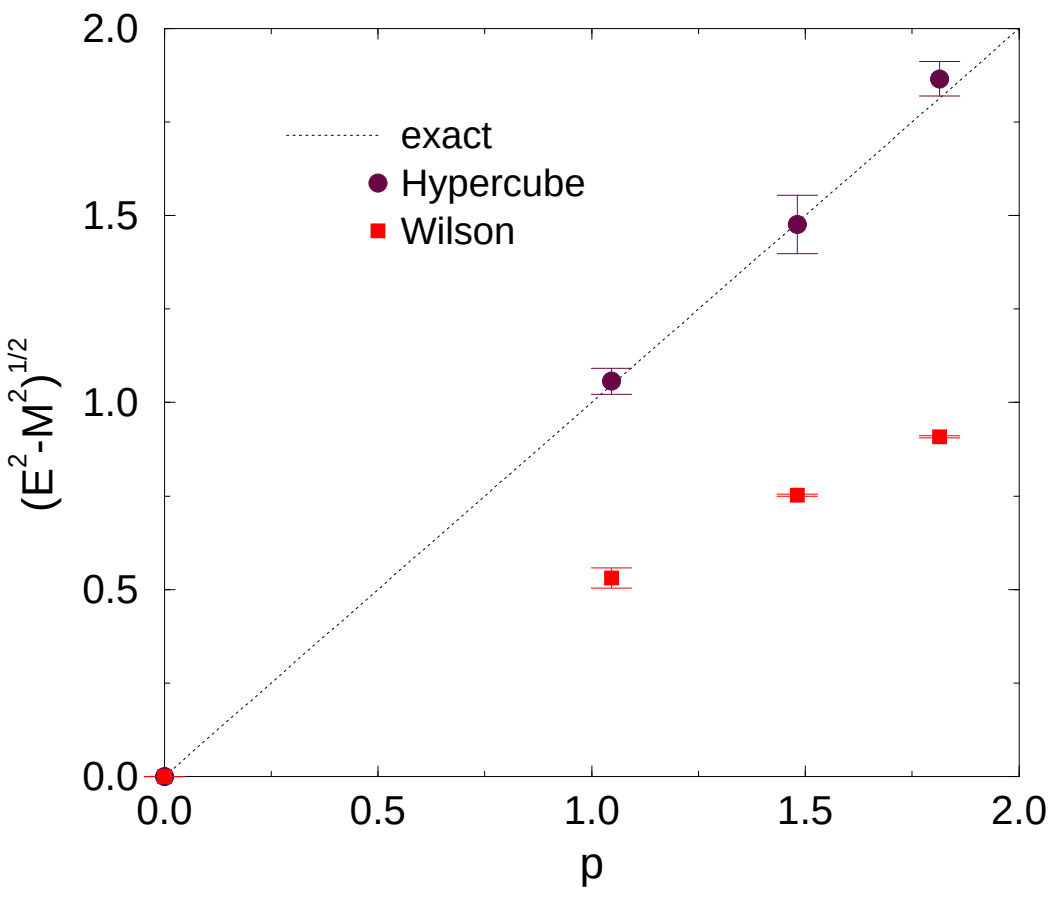

Figure 10. Meson dispersion relation at bare quark mass $m = 0$ in a volume $6^3 \times 18$ at $\beta = 5.0$. The static meson mass $M$ turns out to be 3.0, showing a large renormalization.

Another important quantity for heavy quark physics is the Pauli term $\vec{\sigma} \cdot \vec{B}/2m_B$. For a given lattice action, $m_B$ can be evaluated by introducing a constant magnetic field [10]. The hyperfine splittings depend on this term. When the hypercube action is made gauge invariant by hand, $m_B$ turns out to be clearly improved compared to the Wilson action, but it is still quite bad, see Fig. 11. In the Symanzik approach it is the clover term that is used to obtain the correct Pauli term. In the perfect actions this term arises naturally from the calculation of the perfect quark-gluon vertex. Making the free fermion action gauge invariant by hand is only a crude approximation to a perfect action. Hence there is no reason to be surprised that we did not obtain exact results.

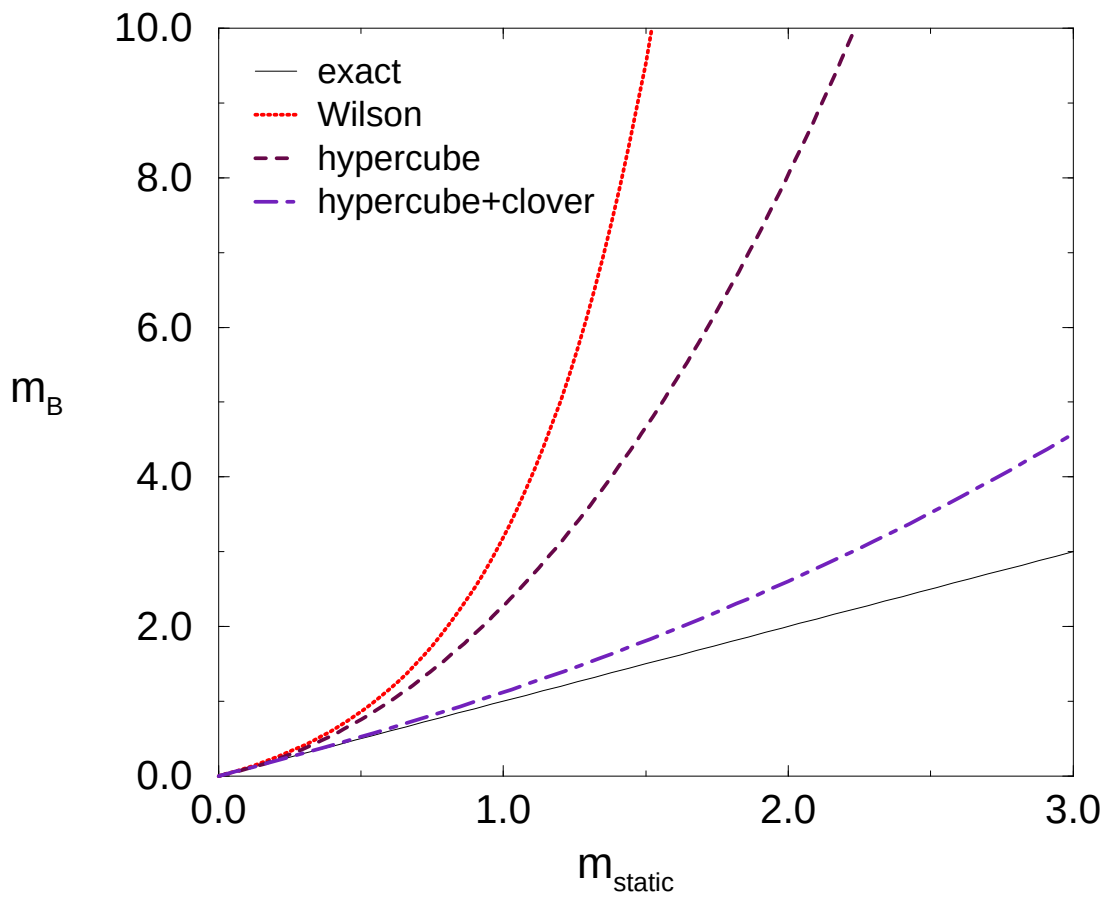

Figure 11. Static mass vs. "magnetic mass" $m_B$.

## 4. The perfect quark-gluon vertex

We now discuss the interacting theory to first order in the quark gluon coupling $g$. Now the blocking from the continuum of a fermionic field involves the continuum gauge field. As an ansatz we write

$$\Psi^i(p) = \sum_{l \in \mathbb{Z}^d} \psi^i(p + 2\pi l)\Pi(p + 2\pi l) + \frac{g}{(2\pi)^d} \sum_l \int d^d q K_\mu(p + 2\pi l, q + 2\pi l) \times a^a_\mu(p - q)\lambda^a_{ij}\psi^j(q + 2\pi l), \quad (14)$$

where $\lambda^a$ are Hermitian generators of the SU(N) algebra. Gauge covariance requires the kernel function $K_\mu$ to obey

$$(p_\mu - q_\mu) K_\mu(p,q) = \Pi(p-q)\Pi(q) - \Pi(p). \quad (15)$$

In the framework of perturbation theory, it is most suitable to determine $K_\mu$ as follows: expand the continuum gauge phase in one cell in terms of $a_\mu$, using boundary conditions that forbid any flux of $a_\mu$ across the surface of the cell. This expansion yields the "boundary condition kernel function"

$$K_\mu^{bc}(p,q) = \frac{(-1)^{d+1}}{p_\mu - q_\mu} \sum_l \frac{l_\mu^2}{l^2} \prod_{\nu=1}^{d} f_{l_\nu}(p_\nu - q_\nu) f_{l_\nu}(q_\nu),$$
$$f_{l_\nu}(p_\nu) = \frac{e^{ip_\nu/2} - (-1)^{l_\nu} e^{-ip_\nu/2}}{p_\nu^2 - (\pi l_\nu)^2}\, p_\nu. \quad (16)$$

An alternative determination, which is more suitable for the non-perturbative search of fixed points by a minimizer that performs inverse RGTs with block factor 2, proceeds as follows: block from the continuum to a lattice of spacing 1 and also to a lattice of spacing 2. Relating the results by a block factor 2 RGT leads to a recursion relation and its iteration to the "recursion kernel"

$$K_\mu^{recu}(p,q) = \frac{1}{4} \sum_{n \geq 0} \frac{\Pi(p)}{\Pi(p/2^n)} \Pi_\mu(\frac{p-q}{2^{n+1}}) \Pi(\frac{q}{2^{n+1}})$$
$$\times \ \sin \frac{q_\mu}{2^{n+1}} \mathcal{K}_\mu(\frac{p}{2^{n+1}}, \frac{q}{2^{n+1}}), \quad (17)$$

where

$$\mathcal{K}_\mu(p,q) = \sum_{\vec{l} \in \mathbb{Z}^{d-1}} \Big[ \prod_{\nu \neq \mu} \cos(p_\nu + \pi l_\nu/2) \times$$
$$\cos(q_\nu - \pi l_\nu/2) ] / [1 + \vec{l}^{\,2} \Big] \ ,$$

and $\vec{l}$ excludes the $\mu$ component. In $d = 1$ the two kernels coincide, since both of them obey the condition (15). In $d > 1$ they are different, but extremely similar. Both have the small momentum expansion

$$K_\mu(p,q) = \frac{q_\mu}{12} \Big[ 1 + \frac{1}{120} \{ p_\mu^2 - 4[p_\mu(p_\mu - q_\mu) + q_\mu^2]$$
$$- \ 5[\vec{p}(\vec{p} - \vec{q}) + \vec{q}^{\,2}] \} + O(\text{momentum}^4) \Big]$$

where $p = (p_\mu, \vec{p}), q = (q_\mu, \vec{q})$. They vanish at the origin, have a peak $< 0.1$ in the region where the momentum components are $O(1)$, and drop quickly to zero at larger momenta. It turns out that the difference of the kernels is really tiny ($O(10^{-5})$) everywhere.

If we apply our method of blocking from the continuum consistently to $O(gA_\mu)$ we obtain the lattice action

$$S[\bar{\Psi}, \Psi, A_\mu] = S[\bar{\Psi}, \Psi] + S[A_\mu] + V[\bar{\Psi}, \Psi, A_\mu] \quad (18)$$

with an interaction term of the form

$$V[\bar{\Psi}, \Psi, A] = \frac{1}{(2\pi)^{2d}} \int_{B^2} d^d p \, d^d q \ \bar{\Psi}^i(-p) \times$$
$$g V_\mu(p,q) A_\mu^a(p-q) \lambda_{ij}^a \Psi^j(q).$$

In momentum space we identified the vertex function analytically [8]:

$$V_\mu(p,q) = \Delta^f(p)^{-1} \Delta_{\mu\nu}^g(p-q)^{-1}$$
$$\times \sum_{l,n \in \mathbb{Z}^d} \frac{\Pi_\nu(p + 2\pi l - q - 2\pi n)}{(p + 2\pi l - q - 2\pi n)^2} (-1)^{l_\nu + n_\nu}$$
$$\times \Big[ \frac{\Pi(p + 2\pi l)}{i(\not{p} + 2\pi \not{l}) + m} i\gamma_\nu \frac{\Pi(q + 2\pi n)}{i(\not{q} + 2\pi \not{n}) + m}$$
$$+ K_\nu(p + 2\pi l, q + 2\pi n) \frac{\Pi(q + 2\pi n)}{i(\not{q} + 2\pi \not{n}) + m}$$
$$+ K_\nu(q + 2\pi n, p + 2\pi l) \frac{\Pi(p + 2\pi l)}{i(\not{p} + 2\pi \not{l}) + m} \Big] \Delta^f(q)^{-1}$$
$$+ \frac{(\widehat{p_\mu - q_\mu}) \Delta_\mu^g(p-q)^{-1}}{\sum_\rho (\widehat{p_\rho - q_\rho})^2 \Delta_\rho^g(p-q)^{-1}}$$
$$\times [\Delta^f(q)^{-1} - \Delta^f(p)^{-1}]. \quad (19)$$

This formula might look a little complicated, but it becomes quite transparent if we consider the "Feynman diagrams" for perfect lattice perturbation theory, illustrated in Fig. 12. They refer to the Landau gauge in the continuum and explain essentially the lengthy term involving $\sum_{l,n}$. Inserting the general lattice gauge propagator $\Delta_{\mu\nu}$ and adding the last two lines takes the vertex function to a general gauge.

The action (18) is gauge invariant to $O(g)$ since $V_\mu(p,q)$ obeys the Ward identity

$$(\widehat{p_\mu - q_\mu}) V_\mu(p,q) = \Delta^f(q)^{-1} - \Delta^f(p)^{-1}. \quad (20)$$

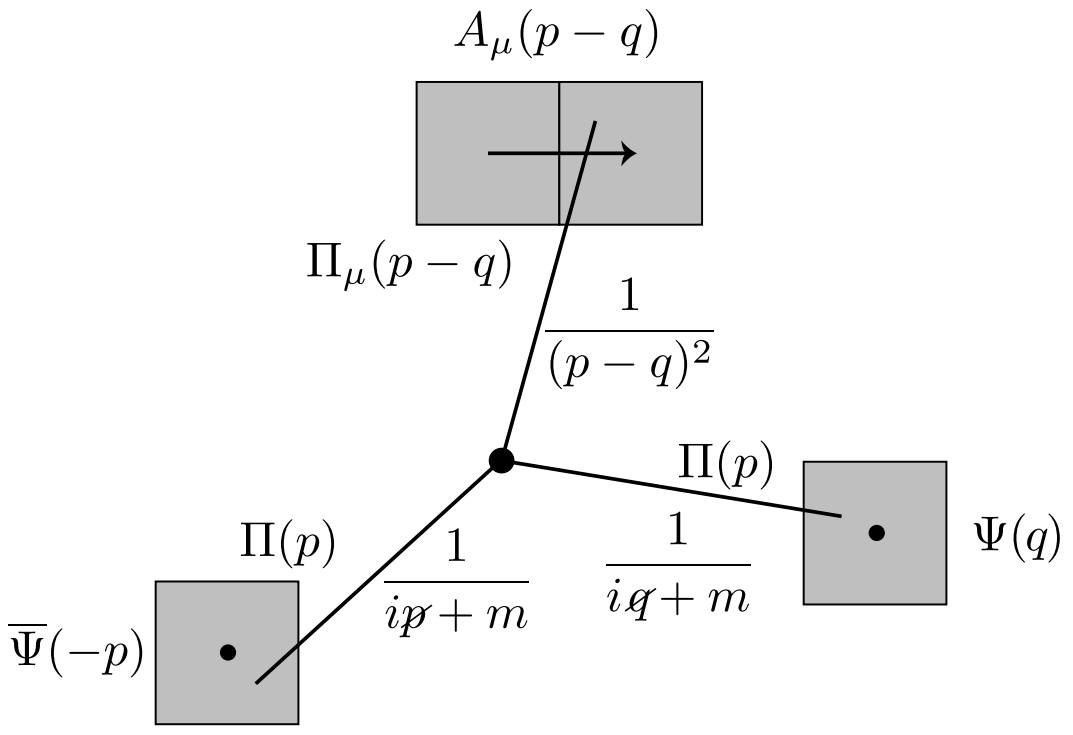

Figure 12. Construction of the perfect quark-gluon vertex function by building blocks which arise from blocking out of the continuum.

To $O(A_\mu)$ this action is free of all lattice artifacts of $O(a^n)$ and $O(ga^n)$, where $a$ is the lattice spacing. Possible artifacts come in at $O(g^2a^n)$.

The construction of the 3-gluon vertex and of higher order terms is straightforward from the same building blocks. Higher order terms include higher and higher dimensional sums in momentum space. They are all UV regular, since we integrate from the beginning over the very short distances (which brings in the $\Pi$ functions). However, it becomes very hard to evaluate such terms and transform them to coordinate space. For the quark-gluon vertex discussed here – with its $2d$-dimensional sum – this is still feasible in $d=4$.

## 5. The perfect clover term in coordinate space

Let us first map the system on $d=2$. To keep the number of parameters small and still observe locality as a crucial property, we even specialize at first on the case where the fermion fields are constant in the 2-direction. Then the inverse Fourier transform of the perfect vertex function can be performed analytically:

$$V[\bar\Psi,\Psi,A_\mu] = \Big(\frac{m}{e^m-1}\Big)^2\Big[\frac{1}{2}\sum_x\Big\{2e^m\bar\Psi_x\Psi_x$$
$$-\bar\Psi_x(1+igA_{1,x+\hat 1/2})\Psi_{x+\hat 1}$$
$$-\Psi_{x+\hat 1}(1-igA_{1,x+\hat 1/2})\Psi_x$$
$$+\bar\Psi_x\sigma_1(1+igA_{1,x+\hat 1/2})\Psi_{x+\hat 1}$$
$$-\bar\Psi_{x+\hat 1}\sigma_1(1-igA_{1,x+\hat 1/2})\Psi_x\Big\}$$
$$+g\sum_{xyz}\Big\{\bar\Psi_x\sigma_2\chi(x_1-y_1,y_1-z_1)iA_{2,y}\Psi_z$$
$$+\bar\Psi\sigma_3\theta(x_1-y_1,y_1-z_1)F_y\Psi_z\Big\}\Big], \qquad (21)$$

$F$ being the gauge field strength defined on the plaquette centers. The interaction terms $\propto 1,\sigma_1$ are obvious by gauge covariance. The interesting parts are the functions $\chi$ and $\theta$. They turn out to be ultralocal, which is also promising for the locality in higher dimensions. The complete table of their non vanishing elements at $m=0$ is given in [8].

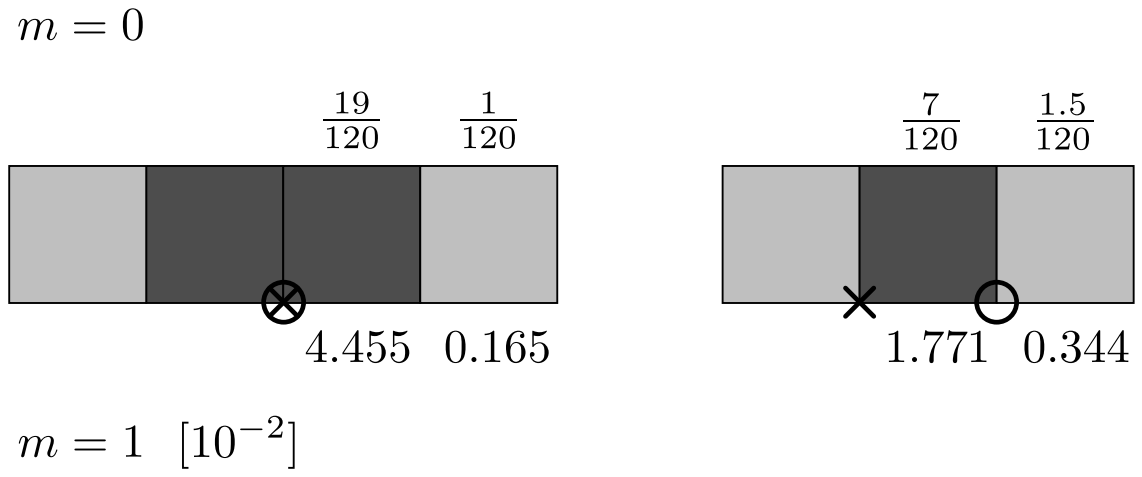

Figure 13. The 2d perfect clover term for fermion fields which are constant in one direction. The symbols $\times$ and $\bigcirc$ mark the positions of $\bar\Psi$ and $\Psi$, and we show the plaquette couplings for $m=0$ (above) and $m=1$ (below, in units $[10^{-2}]$).

In particular $\theta$ determines the "perfect clover term" in this case. We now concentrate on that term $\propto\sigma_3$. An illustration of its plaquette couplings at $m=0$ and 1 is given in Fig. 13. We see that the usual clover term, given by the closest plaquette coupling in the case where $\bar\Psi$, $\Psi$ are at the same position, is indeed dominant. However, there is at least one more contribution which is certainly not negligible. In the case where $\bar\Psi$, $\Psi$ are separated by one link, the coupling to the plaquette attached to this link is suppressed by much less that an order of magnitude. A parameterization in terms of parallel transporters relates this term to the "staple".

Now we come to the general 2d case and impose 3-periodic boundary conditions. Then the plaquette couplings do not follow uniquely from the link couplings – given in formula (19) – any more. If we assume a behavior under unfolding as in the above, effectively 1d case, we obtain the result described in Fig. 14 for $m = 0.001$ and $m = 1$. Again we observe that the "staple term" is only suppressed by about a factor 2 with respect to the usual clover term, whereas all the rest is much smaller.

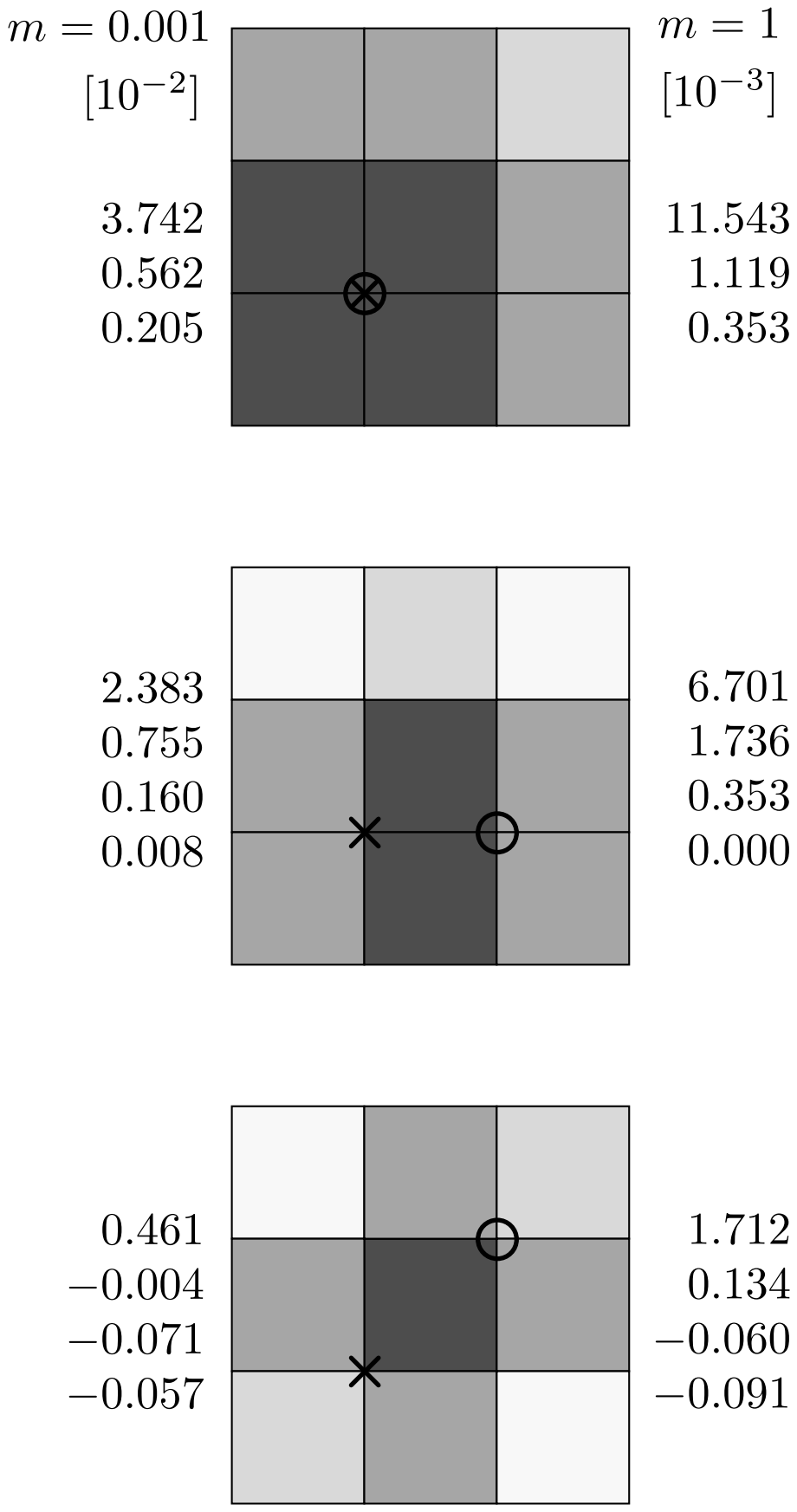

Figure 14. The plaquette couplings in the general 2d perfect clover term after hypercubic truncation. The darkest equivalence class has the largest coupling etc. (at $m = 1$). Again $\times$ and $\bigcirc$ mark the positions of $\bar{\Psi}, \Psi$.

If we sum up all these Pauli terms, i.e. if we consider the case where $\bar{\Psi}$, $\Psi$ are constant everywhere, then we obtain $\frac{1}{2}$ at $m = 0$. This becomes plausible from the identity

$$\partial_\mu \partial_\nu \gamma_\mu \gamma_\nu = \partial_\mu \partial_\mu + \frac{1}{2}[\partial_\mu, \partial_\nu]\sigma_{\mu\nu},$$

which displays this summed Pauli term as a natural partner of the Laplacian. In general, however, the summed Pauli term is a function of the fermion mass, namely

$$s(m) = \Big(\frac{m}{\hat{m}}\Big)^2 \Big[\frac{1}{m} - \frac{1}{\hat{m}}\Big], \quad (\hat{m} = e^m - 1). \qquad (22)$$

This constraint is obeyed by the 2d plaquette couplings given in Figs. 13,14, and the summation of the general 2d case in only one direction also reproduces the result of the effectively 1d case.

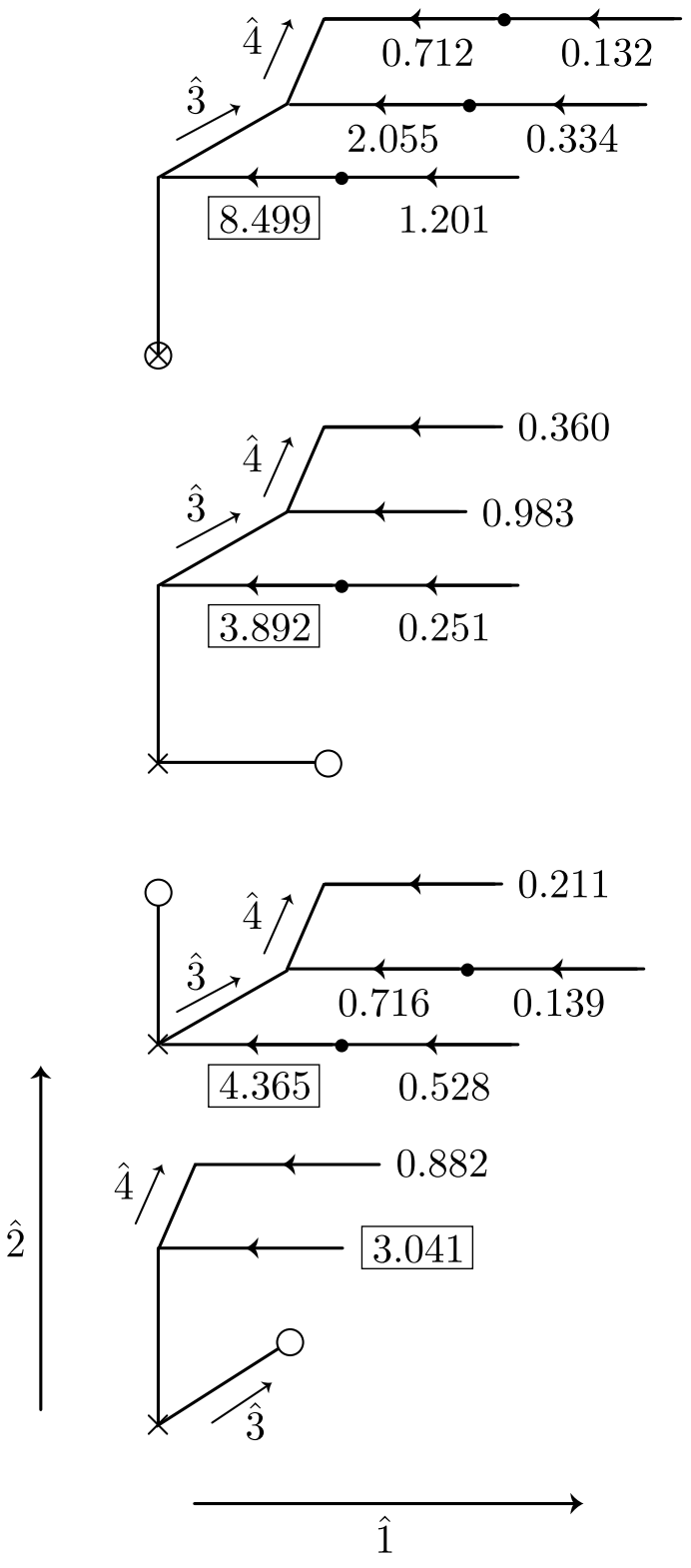

Figure 15. The largest link couplings $\propto \gamma_1\gamma_2$ in the 4d perfect vertex function $V_1$ after hypercubic truncation (in units $[10^{-4}]$). Also here $\times, \bigcirc$ mark the positions of $\bar{\Psi}, \Psi$.

Finally we look at four dimensions. The largest couplings at $m = 1$ are represented in Fig. 15,

where we consider the clover terms $\propto \gamma_1\gamma_2$ in the vertex function $V_1$. Contributions $\propto \gamma_2\gamma_3$ etc. are much smaller, and the rest follows by symmetries. In all the cases where $\bar{\Psi}$, $\Psi$ are separated by diagonals, the couplings are strongly suppressed. If they are separated only by one link, however, we observe in Fig. 15 the presence of important contributions, related to various types of staples. This confirms the trend that such staple terms are only suppressed by a factor $\sim 2$ with respect to the clover leaf. We therefore recommend including at least these extra terms in improved actions.

In spite of the optimization of locality, we have probably still too many couplings for practical applications. It turns out that a crude truncation in coordinate space destroys qualitative properties such as the relation to lower dimensions by setting the remaining momenta equal to 0. So we do not choose that truncation scheme, although its results are quantitatively similar to the ones we show here. The small contributions should not just be omitted but projected onto the surviving couplings. This is exactly what periodicity did for us, but it's capacity is exhausted: a period smaller than 3 annihilates the entire kinetic term of the fixed point fermion and is therefore not sensible.

A simple way to analyze the higher dimensional operators that are introduced by the perfect quark-gluon vertex is an expansion in small momenta. Here we consider the terms in the vertex that have the structure $\sigma_{ij}(p_i - q_i)A_j$, which are the clover like terms. If we assume $A_\mu$ to be constant in Euclidean time ($p_4 = q_4$), the leading term in small spatial momenta has the form,

$$V_j(\vec{p},\vec{q},p_4) = \sigma_{ij}(p_i - q_i)[i\gamma_4(c_1 \sin p_4 + c_2 \sin 2p_4) + d_0 + d_1 \cos p_4 - c_2 \cos 2p_4]. \quad (23)$$

Remarkably, there are no higher order Fourier terms. The coefficients at various masses are given in Table 3. Thus, the perfect vertex suggests a more complicated form for the clover term than is usually used in the lowest order Symanzik improvement program. We already discussed the mass parameter $m_B$ for the hypercube action which was made gauge invariant by hand. If we include the information of the clover terms through couplings $c_1$, $d_0$ and $d_1$, we improve $m_B$ considerably. This improvement is shown in Fig. 11.

| $m$ | $c_1$ | $c_2$ | $d_0$ | $d_1$ |
|---|---|---|---|---|
| 0.01 | -0.219 | -0.0137 | 0.206 | 0.274 |
| 1 | -0.0534 | -0.00287 | 0.0676 | 0.711 |
| 2 | -0.0107 | -0.000425 | 0.0183 | 0.0150 |

Table 3
Coefficients of the clover like terms (described in the text) obtained from the small momentum expansion of the quark-gluon vertex.

## 6. Non-perturbative perfect action

So far we have emphasized our perturbative methods and results. The reasons are two fold: (i) Perturbation theory allows us to get analytical results, whereas the fully non-perturbative RGT with few exceptions (see next Section) must ultimately be formulated numerically as a multigrid minimizer. (ii) Any practical RGT scheme requires truncation in the range of the lattice action so the perturbative results are very useful for assessing the impact of this truncation and giving us a first approximation to the fixed point action. Nonetheless a fully non-perturbative perfect lattice action is the ultimate goal. Here we outline the basic steps and report on the current state of our research.

The (quantum) fixed point action for lattice QCD, $S[U_\mu, \bar{\Psi}, \Psi]$, can in principle be defined by integrating out continuum fields $a_\mu, \bar{\psi}, \psi$ in blocks associated with fixed lattice fields,

$$e^{-S[U,\bar{\Psi},\Psi]} = \int Da_\mu D\psi D\bar{\psi}\; \Delta_{FP}\; \delta(G_{GF}) \times e^{-s[a_\mu,\bar{\psi},\psi] - T_g[U_\mu,a_\mu] - T_f[\bar{\Psi},\Psi,\bar{\psi},\psi]} \quad (24)$$

where $\Delta_{FP}$ is the Faddeev-Popov term and $\delta(G_{GF})$ fixes the gauge. As before, the gauge and fermionic blocking transformation terms, $T_g$ and $T_f$, define the particular RGT by correlating the lattice fields with a local average over the continuum fields. We assume that the fermionic blocking term has been chosen to be quadratic in the quark fields. For the gauge blocking, we must project non-compact continuum fields onto a compact lattice fields, but – as we will report in a separate publication – this requires non-Gaussian terms. The "classically perfect action", which corresponds to the limit $\hbar \to 0$, is found by taking the minimum of the exponent on the right-hand side. The classical limit (or tree approximation for the effective action) greatly simplifies the problem by removing the Faddeev-Popov ghosts and decoupling the quarks from the fixed point gauge action. Thus we are left with an extremum condition for the continuum gauge fields,

$$S_g[U_\mu] = Min_{a_\mu}[\, s_g[a_\mu] + T_g[U_\mu, a_\mu]\,]. \qquad (25)$$

In general we still find it convenient to define the RGT in a fixed gauge internal to each block, so the minimization is done constrained to this gauge surface. Subsequently, the fixed point fermion action is found from the minimum of the quadratic form

$$\bar{\Psi}_y \Delta^{-1}_{y,x} \Psi_x = Min_{\bar{\psi},\psi}[\, s_f[\bar{\psi}, \psi] + T_f[\bar{\Psi}, \Psi, \bar{\psi}, \psi]\,],$$

which can be solved by matrix inversion in the fixed background of the continuum gauge fields $a_\mu$ determined by Eq. 25.

Our approach to this problem is to break it up into two steps. Use perturbation theory to approximate the integrals from the continuum onto a fine lattice followed by a few iterations of the full non-linear RGT on a multigrid, as illustrated in Fig. 16. Let us describe this procedure in some detail for the gauge action. In the classical limit, the perturbative expansion of the perfect action is given as a sum over tree diagrams, with the $O(g^n)$ term contributing uniquely to the $O(n-2)$ monomial in the fields. So far we have computed only the lowest order quadratic contribution – the gluon propagator. From this we can start to approximate the compact lattice action for smooth fields. For example, we have chosen a parameterization with 6 terms, (see Sec.2) which accurately reproduces the quadratic term for loops up to length 8. Each term is represented by a Manton action: e.g. $Tr[Log(U)Log(U^\dagger)]$ where $U$ is the appropriate Wilson loop for each term in the action (26). When we have determined the 3-gluon vertex, a more accurate small field action will be formulated.

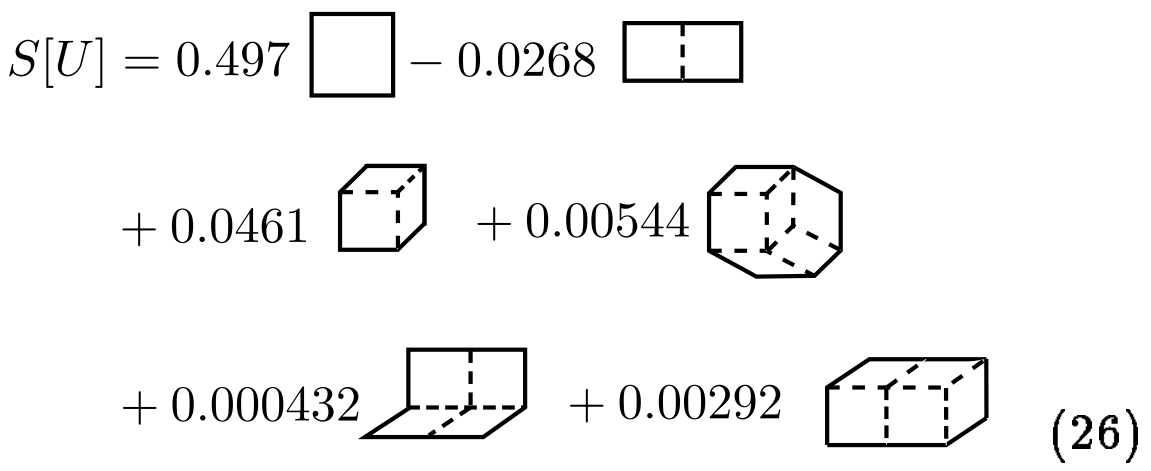

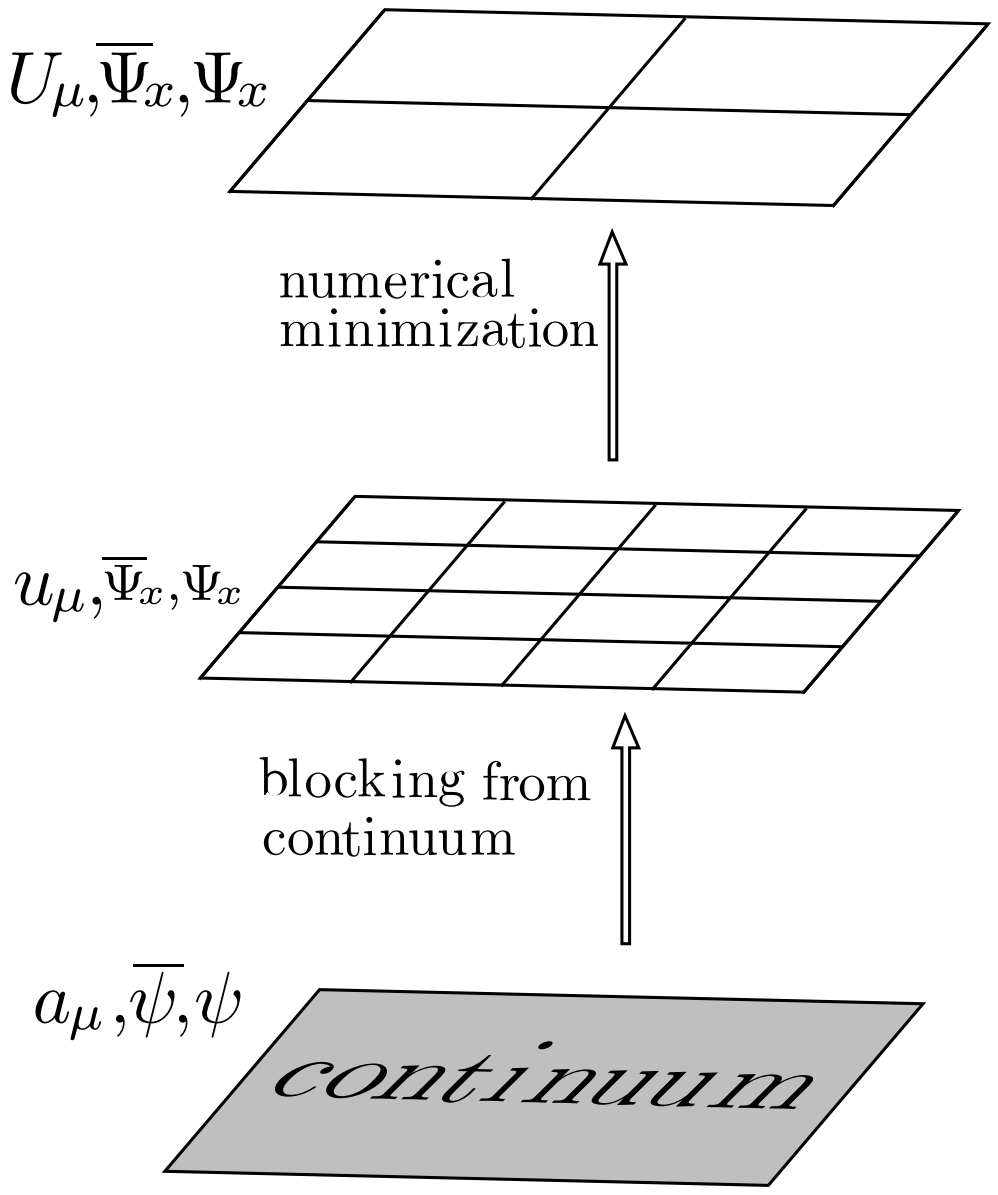

Figure 16. Non-perturbative perfect action by combined continuum and multigrid minimization.

Now we re-express the RGT in Eq. (25) as a recursive blocking by factors of 2. Specifically, we choose to define our blocking transformation in the Landau gauge, expressed in a Manton like form. Each gauge link $U_\mu(x)$ on the doubled lattice is related to the 16 pairs of product links $u_\mu(y)u_\mu(y+\hat{\mu})$ on the original finer lattice that connect the two hypercubes containing $x$ and $x + 2\hat{\mu}$. The gauge blocking function is given by

$$T_g[U_\mu, u_\mu] = \sum_{x,\mu} \sum_{y} Tr[H^\dagger_\mu(x;y) H_\mu(x;y)] \qquad (27)$$

in terms of the Hermitian generators,

$$H_\mu(x;y) = -iLog[U^\dagger_\mu(x) u_\mu(y) u_\mu(y+\hat{\mu})], \qquad (28)$$

for each of the 16 double links. We have written and tested a code for an efficient minimizer for this problem. The multigrid minimizer enforces the gauge constraint,

$$G_x = \sum_{\mu, y \in x} Tr[Log(u_\mu(y)) Log(u_\mu^\dagger(y))] = 0, \quad (29)$$

inside each $2^4$ block characterized by the point $x$, using Lagrange multipliers $\lambda_x$. By applying gradient *descent* in $u_\mu(y)$ combined with a gradient *assent* in $\lambda_x$ to

$$s_g[u_\mu] + T[U_\mu, u_\mu] + \sum_x \lambda_x G_x,$$

we obtain an efficient algorithm. Fortunately the gauge constraint does not present any essential difficulty to the formulation or application of the algorithm. Also the Manton form of the blocking term (27), since it is zero at its minimum, has the advantage that there is no need to subtract a field dependent factor to normalize the FPA. Our hope is that by an accurate perturbative starting point, the fixed point can be found with one or two additional iteration steps. Moreover it is also possible within this general framework to simulate RGTs at finite correlation length by the MCRG to obtain quantum corrections to the classically perfect trajectory. In the next section we turn to a simple example where classical and quantum perfection actions can be computed analytically and compared.

## 7. Quantum perfect vs. classically perfect topological charge

In this Section we work out the concept of a quantum perfect (really perfect) topological charge and study the artifacts of the classically perfect charge [13]. As an example we consider the 1d XY model, which is equivalent to a quantum mechanical particle on a circle. Its propagator from angle 0 at Euclidean time 0 to $\phi$ at time $t$ can be written as a path integral

$$\langle \phi, t | 0, 0 \rangle_\theta = \int_{(0,0)\to(\phi,t)} D\varphi(\tau) e^{-S_\theta[\varphi(\tau)]} \quad (30)$$

$$= \sum_{Q \in \mathbb{Z}} \int_{(0,0)\to(\phi,t)}^{\varphi(\tau)\in\omega_Q} D\varphi(\tau) \exp\{-S[\varphi(\tau)] + iQ\theta\}$$

where $\tau \in [0, t]$, $\omega_Q$ is the set of paths with the given endpoints and winding number $Q$, and we have included a $\theta$ term. For fixed $\omega_Q$ we can insert the free propagator on $\mathbb{R}$ and obtain the Villain form

$$\langle \phi, t | 0, 0 \rangle_\theta = \sqrt{\frac{I}{2\pi t}} \times \quad (31)$$
$$\sum_{Q \in \mathbb{Z}} \exp\{-\frac{I}{2t}(\phi + 2\pi Q)^2 + iQ\theta\}.$$

where $I$ is the moment of inertia. Of course we can insert discrete points between 0 and $t$ and integrate over all possible intermediate states (Chapman-Kolmogoroff equation). For example, if we consider $N$ equidistant points and insert the exact transfer matrix, then we obtain the perfect lattice description,

$$\langle \phi_N, t | \phi_0, 0 \rangle_\theta = \Big(\frac{I}{2\pi\Delta t}\Big)^{N/2} \sum_{n_N} \prod_{j=1}^{N-1} \sum_{n_j} \int_{-\pi}^{\pi} d\phi_j$$

$$\exp\{-\frac{I}{2\Delta t}[(\phi_N - \phi_{N-1} + 2\pi n_N)^2 + \ldots$$
$$+(\phi_1 - \phi_0 + 2\pi n_1)^2] + iQ\theta\}, \quad (32)$$

where $Q = n_1 + n_2 + \ldots + n_N$ and $\Delta t = t/N$. $n_j$ is the winding number between the positions $\phi_{j-1}$ and $\phi_j$. The perfect action shown here is an RGT fixed point, since the 1d FPA of a free scalar particle can be made ultralocal by suitable parameters. However, this description not only consists of a perfect action but also of a prescription for the measure. We do not attach one particular charge to a given lattice configuration, as it is usually done by some smooth interpolation. Instead we integrate over all possible interpolations of the lattice configuration, including all winding numbers $n_j \in \mathbb{Z}$ in each discrete step. This is the very idea of path integration. Now the topological charge is given by a probability distributions $p(\{\phi_j\}, Q)$ for each lattice field configuration $\{\phi_j\}$. Thus it can never happen that "small windings fall through the lattice meshes".

This is in contrast to the classically perfect action, where some topological charge gets lost in this way. This was observed to be significant on very coarse lattices in the 2d O(3) model, CP(3) model and pure SU(2) gauge theory [14]. In our case the classically perfect charge corresponds to the geometrical definition of the topological charge, i.e., neighboring lattice points are interpolated by the shortest arc. Hence $n_j$ is confined to $\{-1, 0, 1\}$. Fig. 17 illustrates the artifacts of the classically perfect action.

The energy eigenvalues of the 1d rotor are

$$E_\ell(\theta) = \frac{1}{2I}\left(\ell + \frac{\theta}{2\pi}\right)^2, \qquad \ell = 0, 1, 2, \ldots$$

and the correlation time is identified as

$$\xi = \frac{1}{E_1(0) - E_0(0)} = 2I.$$

The perfect topological susceptibility is given by

$$\chi = \frac{1}{t}\langle Q^2 \rangle = -\frac{1}{t}\partial_\theta^2 Z_t(\theta),$$

where $Z_t(\theta) = \langle 0, t | 0, 0 \rangle_\theta$ is the partition function. Perfect scaling requires the product $\chi\xi$ to be constant. Fig. 17 shows the scaling artifacts in this quantity for the standard and for the classically perfect action.

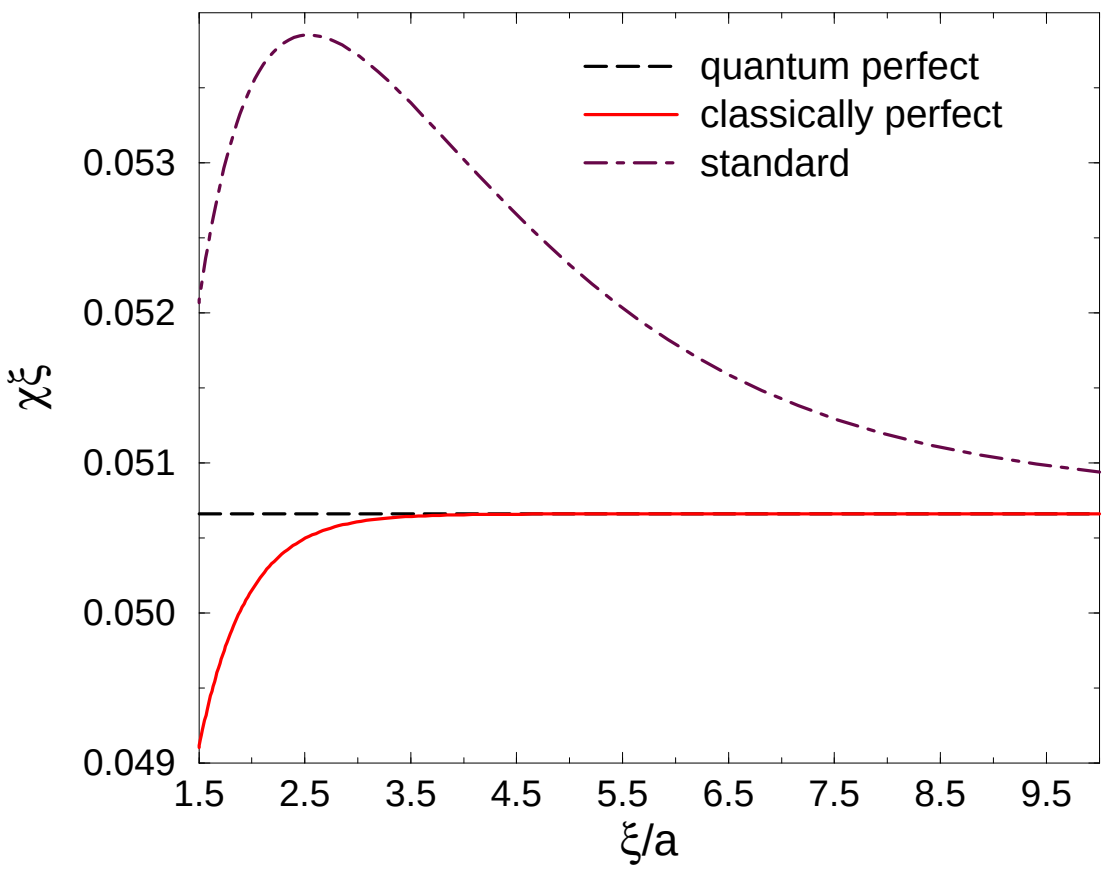

Figure 17. Artifacts in the scaling quantity $\chi\xi$ as function of the correlation time $\xi$ in lattice units.

The Villain type action is generally encountered when one considers exact projections from continuum fields onto compact lattice fields along the lines discussed in Sec. 6.

To *summarize*, we have outlined our program for constructing perfect actions for lattice QCD in the classical approximation. The results obtained from perfect lattice perturbation theory are suitable for applications to heavy quark physics. Moreover they provide a basis for a multigrid construction of non-perturbative perfect actions, suited to strongly fluctuating fields.